# Emergent Bound States and Impurity Pairs in Chemically Doped Shastry-Sutherland System

Zhenzhong Shi,[1] William Steinhardt,[1] David Graf,[2] Philippe Corboz,[3] Franziska Weickert,[2] Neil Harrison,[4] Marcelo Jaime,[4] Casey Marjerrison,[1] Hanna Dabkowska,[5] Frédéric Mila,[6] Sara Haravifard[1,7*]

[1]Department of Physics, Duke University, Durham, North Carolina 27708, USA

[2]National High Magnetic Field Laboratory, Florida State University, Tallahassee, Florida 32310, USA

[3]Institute for Theoretical Physics and Delta Institute for Theoretical Physics, University of Amsterdam, Science Park 904, 1098 XH Amsterdam, The Netherlands

[4]National High Magnetic Field Laboratory, Los Alamos National Laboratory, Los Alamos, New Mexico 87545, USA

[5]Brockhouse Institute for Material Research, McMaster University, Hamilton, Ontario L8S 4M1, Canada

[6]Institut de Théorie des Phénomènes Physiques, École Polytechnique Fédérale de Lausanne (EPFL), CH-1015 Lausanne, Switzerland

[7]Department of Mechanical Engineering and Materials Science, Duke University, Durham, North Carolina 27708, USA

[*]Corresponding author (email: haravifard@phy.duke.edu)

**The search for novel unconventional superconductors is a central topic of modern condensed matter physics. Similar to other Mott insulators, Shastry-Sutherland (SSL) systems are predicted to become superconducting when chemically doped. This makes $SrCu_2(BO_3)_2$, an**

**experimental realization of SSL model, a suitable candidate and understanding of the doping effects in it very important. Here we report doping-induced emergent states in Mg-doped SrCu$_2$(BO$_3$)$_2$, which remain stable up to high magnetic fields. Using four complementary magnetometry techniques and theoretical simulations, a rich impurity-induced phenomenology at high fields is discovered. The results demonstrate a rare example in which even a small doping concentration interacts strongly with both triplets and bound states of triplets, and thus plays a significant role in the magnetization process even at high magnetic fields. Moreover, our findings of the emergence of the very stable impurity pairs provide insights into the anticipated unconventional superconductivity in SrCu$_2$(BO$_3$)$_2$ and related materials.**

Geometrical frustration in low-dimensional quantum spin systems often leads to exotic states of matter where new physics can emerge[1]. However, understanding these systems is often hindered by the complex Hamiltonians in analytical and numerical studies. SrCu$_2$(BO$_3$)$_2$, a realization of the exactly solvable SSL model[2], provides an important test ground for our understanding of frustration in quantum magnets. It consists of two-dimensional layers of Cu$^{2+}$($S$ = 1/2) orthogonal dimers arranged on a square lattice (Fig. 1a inset). A spin gap $\Delta \sim 3$ meV separates the $S$=0 singlet ground state from an $S$=1 triplet excited state. As a spin-gapped Mott insulator[3], SrCu$_2$(BO$_3$)$_2$ has been proposed to host a resonating valence bond (RVB) type superconductivity upon doping. Doping-induced superconductivity is suggested to form either by introducing enough mobile charge carriers into the CuBO$_3$ plane[3–7], or by inducing a small distortion in the lattice so that the system selects an energetically-nearby state where charge carriers can move freely[8]. Experimental attempts continue and efforts are still ongoing to find such evidence of superconductivity in $SrCu_2(BO_3)_2$[9–14].

To proceed, a central question is to understand the ground state induced by the interplay among the non-magnetic spin singlets, the magnetic spin triplet excitations, and the doping-introduced $S = 1/2$ spin singlet impurities at low energies. A suitable candidate for such a study is $SrCu_{2-x}Mg_x(BO_3)_2$, in which the magnetic $Cu^{2+}$ is substituted with non-magnetic isoelectronic $Mg^{2+}$, introducing minimal structure distortion because of their similar ionic radii. Previous inelastic neutron scattering and $\mu$SR experiments on $SrCu_{2-x}Mg_x(BO_3)_2$(x=0.05) have shown that some dimers are indeed broken, and in-gap states emerge[13,14]. It was suggested that the in-gap states might correspond to localized anisotropic spin polarons developed around the impurities[13,15], or to the $S = 1/2$ states that consist of one spinon and one impurity[16]. However, to this date a clear understanding for the effects of non-magnetic impurities in SSL systems remains elusive. Here we demonstrate that critical insights are gained by studying the magnetization response of the chemically doped $SrCu_2(BO_3)_2$ in high magnetic fields.

In the presence of magnetic field, frustration has been known to induce magnetization plateaus, occurring at fractional values of saturation magnetization $M_{sat}$, either due to a "classical" mechanism involving stabilization of some classical spin configurations, or due to a "quantum" mechanism which corresponds to symmetry-breaking phase transitions in an effective hard-core-boson model[1]. As one of the best examples of the latter case, $SrCu_2(BO_3)_2$ exhibits a series of magnetization plateaus at magnetic fields above which the spin gap is closed by the Zeeman energy[17–22]. This has been understood as a result of the crystallization of $S_z = 2$ "pinwheels" of bound states of two triplets[23], which are energetically more favorable than crystals of $S_z$ = 1 triplets. This picture is well established in the pure system. In the doped system, however, it is not clear how the added impurities would interact with the triplets and bound states of triplets, and hence alter their crystallization. Here, we report a comprehensive doping dependence study of the magnetometry in high magnetic fields, revealing a surprisingly rich impurityinduced phenomenology in these systems: doping-induced triplet states and emergent impurity pairs. It was found that the conventional magnetization measurements alone do not provide a full picture explaining the subtle changes associated with such a study. Therefore, we combined four complementary techniques: tunnel diode oscillator (TDO) and torque magnetometry which measure magnetic susceptibility; magnetization measurements which probe magnetization response directly; magnetostriction measurements which detect lattice correlations to the magnetic order in very high magnetic fields. The results were confirmed by our numerical simulations using infinite projected entangled pair states (iPEPS), providing an accurate account of the various impurity-induced emergent states. Our results offer essential implications for the understanding of doped quantum spin systems and consequently aid us in our quest for new unconventional superconductors.

## Experimental Results

The tunnel diode oscillator (TDO) and torque magnetometry experiments were conducted at the National High Magnetic Field Laboratory (NHMFL) dc field facility, while the magnetization and magnetostriction measurements were carried out at the NHMFL pulsed field facility. Single crystals of $SrCu_2(BO_3)_2$ and $SrCu_{2-x}Mg_x(BO_3)_2$, with $x$ up to 0.05, were grown using the optical floating zone technique (see Methods). In the doped samples, the magnetic $Cu^{2+}$ sites are replaced with non-magnetic isoelectronic $Mg^{2+}$, which effectively breaks the spin dimers into free $S = 1/2$ spins, without introducing structural distortions. The doping concentrations were confirmed by the susceptibility measurements (see Fig. S1).

## Magnetization plateaus

We show in Fig. 1a the magnetization response for SrCu$_{2-x}$Mg$_x$(BO$_3$)$_2$ ($x$ = 0.02, 0.03, and 0.05) at 0.4 K with $H \parallel a$ up to 60 T. At low fields, a notable finite magnetization, which increases with doping, is observed (see Fig. 1a lower right inset). For $H$ smaller than 6 ~ 8 T, $M(H)$ exhibits a Brillouin-like paramagnetic behavior for all three dopings, and the results are consistent with the field-induced alignment of free $S = 1/2$ impurity spins. However, a full saturation of magnetization is interrupted at $H$ above 6 ~ 8 T, suggesting a more complicated picture than one soley explained by the impurity-induced free spins; as will be discussed later. The sharp onset of magnetization at $H$ higher than ~18 T is attributed to the increase in population of triplets, as spin gap closes with increasing field[1]. In case of doped SrCu$_2$(BO$_3$)$_2$, however, the magnetization is suppressed with increasing doping, suggesting a suppressed density of triplets in the presence of impurities. For all doping concentrations, we can extract a reference magnetization $M_{ref}$= 3×$M_{1/3}$, where $M_{1/3}$ refers to the magnetization at the 1/3 plateau. It is noted that for the pure sample, $M_{ref}$ is equal to the saturation magnetization or $M_{sat}$, with all the magnetic moments fully saturated. For the doped samples $M_{ref}$ is expected to be only slightly larger than the saturation magnetization, attributed to the overvaluation of the magnetization of the free spins located on broken dimers with impurity sites (up to 2.5% with the highest doping $x$ = 0.05). As shown in Fig. 1a, however, experimental finding is surprising and $M_{ref}$ in fact decreases as doping concentration, $x$, increases. We plot in Fig. 2 the normalized magnetization curves, $M/M_{ref}$, as function of $x$. More interestingly, $M/M_{ref}$ curves for the pure[24] and doped samples overlap and show the same sequence for plateaus at 1/8, 1/4, and 1/3, which become increasingly softened with doping. This observation is quite unexpected, and suggests that the pure and doped systems may share the same underlying spin structures[23] at these plateaus, enabled by some collaborative geometrical arrangement of the impurities and the triplets in the doped samples.

Complementary magnetostriction measurements (see Methods) performed for both the pure and doped samples are plotted in Figs. 1b and S2. These results reveal contraction along the *a*-axis, which closely corresponds to changes in magnetization and is consistent with previous results reported for the pure system[20,25]. Furthermore, these results clearly show the 1/8, 1/4, and 1/3 plateaus, for which the onset fields agree very well with those determined from the magnetization measurements. The increasingly softened plateaus with doping, also suggest that the overall lattice coupling is suppressed with the increased density of impurities.

## Emergent magnetization states at low *H*

Our most remarkable results are obtained from a close examination of the magnetization response for the doped samples in the field region below the 1/8 plateau, as presented in Figs. 3 and 4. In fact this region is of broad interest, though not well understood even in the pure system. For example, other than the 1/9 plateau[19,20], spin superstructures with even smaller fractions, i.e. larger unit cells, remain elusive. Theoretical considerations seem to suggest that they are energetically favorable only in very limited field ranges, if at all possible[23]. In a doped system, the phase diagram becomes even richer as the density of impurities increases.

We plot in Figs. 3a and 3b the magnetization curves as function of *x* for the low-field region. Indeed, the magnetization measurements clearly show three low-field anomalies, i.e. jumps in $M(H)$ curves, as can be seen for all doped samples. For the $x = 0.05$ sample, the onset fields of these anomalies are determined as $H'_{C1} \sim 17.1$ T, $H'_{C2} \sim 21.7$ T, and $H'_{C3} \sim 25.0$ T (see Figs. S3a and S4). For the $x = 0.02$ and $0.03$ samples, three anomalies are also identified at similar fields. Fig. 3b shows the magnitudes of these anomalies, measured by $dM/dH$, are much smaller than that of the 1/8 plateau for all $x$. Their doping dependence, however, are exactly the opposite: the $H'_{C1}$, $H'_{C2}$, and $H'_{C3}$ anomalies are enhanced with higher doping concentration, while the 1/8 plateau is suppressed, suggesting their different origins. The broad maxima at very low $H \sim 1$ T – 2 T are attributed to field-aligned free $S = 1/2$ impurity spins, associated with the onset of finite magnetization, as discussed above.

We plot in Figs. 3c and 3d the results for TDO magnetic susceptibility measurements (see Methods), where $df/dH$ is proportional to $dM^2/d^2H$; the corresponding comparison is clearly shown in Fig. 3d for the $x = 0.05$ sample. The TDO measurements performed in a quieter magnet environment, i.e. a steady magnetic field instead of a pulsed field, show more clearly the emergence of the $H'_{C1}$, $H'_{C2}$, and $H'_{C3}$ anomalies with doping, and their absence in the pure system (see Fig. 3c). Strikingly, another broad anomaly at $H'_{C0} \sim 9$ T, which is much weaker than its higher field counterparts, is only observed for the highest doping concentration $x = 0.05$ sample. The confirmation of such a weak anomaly underlines the importance of adopting different techniques for measuring the same physical quantity when the signal is weak. The temperature dependence of the anomalies (see Fig. 3d inset, Figs. S5 and S6) shows that they persist up to fairly high temperature of $T \sim 2$ K.

The coupling of these anomalies to the lattice is investigated using magnetostriction measurements. In both pure and doped samples, the axial magnetostriction along a-axis deviates from zero at fields that gradually decrease from ∼ 18 T in $x$ = 0 to ∼ 14 T in the $x$ = 0.05 sample, as indicated by arrows in Fig. 4a. As can be seen in Fig. 4b, however, no anomalies are observed for any of the samples at fields below the 1/8 plateau. Though a lack of sufficient resolution cannot be completely ruled out, the absence of these anomalies in the magnetostriction data suggests their weak coupling to lattice. This is unlike the strong lattice coupling observed for magnetization plateaus corresponding to the crystallization of bound states of triplets, indicating their different origins. This interpretation is further strengthened by the fact that the $H'_{C0}$ and $H'_{C1}$ anomalies appear at fields comparable with, or below, the gap closing fields at which bound states of triplets are absent. Indeed, iPEPS numerical simulations clearly demonstrate that the observed anomalies all have impurity-induced origins, as we explain in the following section.

## Simulation results

### Infinite projected entangled pair states

Our simulation results are obtained using infinite projected entangled pair states (iPEPS) - a variational tensor network ansatz to represent a 2D ground state directly in the thermodynamic limit[26–28]. The ansatz consists of a unit cell of tensors which is periodically repeated on the infinite lattice, where in the present case we use one tensor per dimer[23,29,38]. The accuracy of the ansatz can be systematically controlled by the bond dimension $D$ of the tensors.

The optimization of the variational parameters has been done using the simple update method which provides good estimates of ground state energies while being computationally affordable, even in the limit of very large unit cell sizes (up to 12 × 12 dimers in the present work). For the computation of observables, a variant[30,31] of the corner-transfer matrix method[32,33] is used. To improve the efficiency, we exploit the U(1) symmetry of the model[34,35]. For an introduction to the method, see Refs. 36,37 for example.

## Model used for the Mg-doped SrCu2(BO3)2

A well-established effective model to describe the low-energy physics of SrCu$_2$(BO$_3$)$_2$ is the SSL model[2] given by the Hamiltonian

$$H = J \sum_{\langle i,j \rangle} S_i \cdot S_j + J' \sum_{\langle i,j \rangle} S_i \cdot S_j - h \sum_i S_i^z \qquad (1)$$

where the bonds with coupling strength $J$ build an array of orthogonal dimers and the bonds with coupling $J'$ denote inter-dimer couplings, and $h$ is the strength of the external magnetic field. In the present work we use $J'/J = 0.63$ (with $J \sim 51$ T) which was obtained from a fit to high magnetic field data[38].

At zero external magnetic field, the ground state is given by a product of singlets on the dimers[2]. Early on, it was found that the SSL model has almost localized triplet excitations[39,40] which has led to the viewpoint that the magnetization plateaus found in SrCu$_2$(BO$_3$)$_2$ correspond to crystals of triplets[39,41–51]. However, it was predicted that $S_z = 2$ excitations, which can be seen as a bound state of two triplets, are energetically lower in the dilute limit of excitations[42]. Based on iPEPS simulations, it was shown that these bound states are energetically favored even when they are localized, i.e. that the magnetization plateaus actually correspond to crystals of localized bound states rather than crystals of triplets[23].

We model the Mg doping by introducing "impurity" sites, where each impurity replaces one of the $S = 1/2$ spins on a dimer with a non-magnetic site (i.e. with no coupling to the neighboring sites), leaving a free $S = 1/2$ spin on the other site of the dimer. A single impurity in the lattice leads to a two-fold degenerate ground state, since it costs no energy to flip a single spin. Thus in the dilute limit of impurities we can expect that these free $S =$ 1/2 spins immediately align with an external magnetic field. The question is now how the presence of these impurities with attached $S = 1/2$ spins affects the magnetization process, which we will investigate below using iPEPS simulations.

## iPEPS simulations results

To understand the impurity effects in the doped samples, the first key question is whether the bound states of triplets are effectively attracted or repelled by an impurity site and its neighboring $S = 1/2$ spin. To answer this question, we have performed simulations with a single impurity and one bound state in an 8×8 unit cell using a bond dimension $D = 10$, and found that the latter is clearly repelled by the impurity (see Fig. S7). Thus, based on this result we can expect that in a large system containing many impurities, bound states are first created far away from neighboring impurities as magnetic field increases.

Figure 5a shows the iPEPS magnetization curve ($D = 6$) obtained using a 12 × 12 unit cell of dimers with a random configuration of 8 impurity sites, corresponding to a doping $x = 0.056$. For this system, a localized bound state first occurs at $H'_{C2} = 0.428J$ (∼ 21.8 T) which is close to the critical field to create a localized bound state without impurities, $0.427J$ (Ref. 23). For an infinite system, since it will contain many locations in the lattice with similar energy costs to form a bound state, we may expect that the magnetization curve is not smooth but that it exhibits a jump at $H'_{C2}$, compatible with the anomaly observed in experiments at $H'_{C2} \sim 21.7$ T (see Fig. 3).

Upon further increasing the magnetic field, the lattice gets occupied by more and more localized bound states. At a certain characteristic field $H'_{C3} = 0.454J$ (∼ 23.2 T) we observe a change of slope in the magnetization curve, starting from which the lattice is also populated by additional triplet excitations. This can be understood from the fact that a bound state occupies more space than a triplet excitation, so that at locations with several nearby impurity sites it can become energetically favorable to place a triplet excitation rather than a bound state. Thus, this suggests that the anomaly observed at $H'_{C3}$ ~ 25 T in experiments is due to a change of slope in the magnetization curve. An example spin configuration at $H'_{C3}$ is presented in Fig. 5c, containing 3 bound states and one triplet excitation. A good qualitative agreement between the simulation and experiments is demonstrated in Fig. 5b, which shows the doping dependence of the normalized magnetization $M/M_{ref}$ at $H'_{C2}$ and $H'_{C3}$, extracted from the magnetization measurements (see Fig. 3) for the $x$ = 0.02, 0.03 and 0.05 samples and the iPEPS simulations of $M/M_{sat}$ for $x$ = 0.056.

Finally, we address the additional features at $H'_{C1}$ ~ 17.1 T and $H'_{C0}$ ~ 9 T observed in experiments. As explained above, in the dilute limit of impurities we expect all the attached $S$ = 1/2 moments to be aligned already at a small magnetic field. However, at larger doping there is an increasing probability of having two neighboring impurities, as shown in Fig. 5d. In this configuration, the $S$ = 1/2 spins attached to the impurities can no longer be regarded as free, but they prefer to couple to a singlet. As a consequence, the two $S$ = 1/2 spins do not immediately align with a small external field, but only do so beyond a certain critical field. From computing the excitation energy in a 8×8 cell we find a critical field $H'_{C1}$ = 0.238J (~12.1 T), i.e. well below $H'_{C2}$. This value corresponds to the excitation energy in the limit of an isolated pair of neighboring impurities. In the presence of additional nearby impurities (e.g. a third impurity with an attached aligned spin in the vicinity of the impurity pair) the excitation energy will be higher, leading to a collaborative arrangement of impurities and additional energy excitation levels in between $H'_{C1}$ and $H'_{C2}$. As shown in Fig. 5e, there exist also other two-impurity configurations at lower excitation energies, which is consistent with the experimental observation of the broad maximum at an onset field $H'_{C0} \sim 9$ T.

## Discussion

One of the most prominent properties of SrCu$_2$(BO$_3$)$_2$ is the sequence of magnetization plateaus at 1/8, 1/4, and 1/3 of the saturation magnetization, which have been shown to correspond to various superstructures that break the translational symmetry of the lattice. It is remarkable that the high field magnetization curve exhibits even more features in the presence of impurities. Impurities create local defects that are usually saturated by a small field. Our findings, however, portray a very different picture. Here, even a small concentration of impurity plays an important role in the magnetization process at very high magnetic fields. It highlights a nontrivial interplay between impurities and the triplets and bound states of triplets.

For example, the emergence of the $H'_{C2}$ and $H'_{C3}$ anomalies upon doping is unexpected and striking. Even though the localized bound states of triplets are suggested to appear at similar fields in the pure system[23], there is no anomaly observed in magnetization curve. Only in the presence of impurities, the $H'_{C2}$ and $H'_{C3}$ anomalies are stabilized. We believe this is because while in the pure system delocalized bound states appear at $H'_{C2}$, leading to smoothly increasing magnetization curve (i.e. no anomaly), in the doped system the bound states cannot delocalize anymore due to the presence of the impurities. At $H'_{C2}$ all the locations which are sufficiently far away from the impurities (and which have a similar energy cost to form a bound state) will be populated by a bound state, leading to a small jump in the magnetization (i.e. an anomaly) at $H'_{C2}$. The anomaly at $H'_{C3}$ associated with the appearance of additional triplet excitations is not present in the pure case, because in that case the magnetization plateaus correspond to regular crystals of bound states (i.e.

triplet excitations are absent). It is interesting to speculate how these states evolve, and if more exotic states with some special configurations of impurities and triplets would appear at higher doping.

The observation of the impurity pairs that survive up to $H'_{C0}$ and $H'_{C1}$ is another significant result. With a binding energy $\gtrsim g\mu_B H'_{C0} \sim 1$ meV, these spin-singlet impurity pairs seem to be just one step away from becoming Cooper pairs, which require holes (or electrons) to be associated with the impurities. If the energy cost for such doping mechanism is smaller than the binding energy, Cooper pairs could in principle exist. Note that the impurity pairs that we observe are strongly localized, suggesting that Cooper pairs, if exist, might also be strongly localized. Therefore, our results indicate that searches for superconductivity in charge doped $SrCu_2(BO_3)_2$ might benefit from local experimental probes, such as scanning tunneling microscopy (STM) or nuclear quadrupole resonance spectroscopy (NQR). Previous studies on transport and magnetization properties of the charge-doped $SrCu_2(BO_3)_2$ have found no evidence of global superconductivity[11], though there have been no studies reported using local probes on the same samples. It is noted that, in cuprates, as well as the doped $SrCu_2(BO_3)_2$ studied in the Ref. 11, the impurities reside in between the copper-oxide planes, while in $SrCu_{2-x}Mg_x(BO_3)_2$, $Mg^{2+}$ replaces $Cu^{2+}$ in the $CuO_4$ plane. Further studies are required to explore the difference between the impurity configurations in the two cases. Nevertheless, it is not unreasonable to speculate that localized impurity pairs can still be relevant in samples studied by Ref. 11, in which the spin dimers are broken by doping. If Cooper pairs can indeed be induced in this manner, it is likely that with high enough doping, the charge hopping process might be strong enough to support a globally-coherent RVB-type of superconducting phase, as predicted by various theoretical studies[3–8].

Doping with nonmagnetic static impurities is also expected to suppress $M_{sat}$, and disrupt the superstructures. The inferred $M_{ref}$ is indeed suppressed for the doped samples, though by a more-than-expected amount (see Fig. 1a). The $x = 0.05$ sample, for instance, has a loss of 2.5% of magnetic $Cu^{2+}$ moments compared to the pure sample, but its $M_{ref}$ is ~ 17% less than $M_{sat}$ of a pure sample. This can be readily understood as the softening of the superstructures underlying the plateaus by the added impurities: the formation of the 1/3 superstructure is perturbed in a certain neighborhood of the impurities, resulting in patches of the superstructure rather than a perfect 1/3 superstructure with 2.5% of the sites removed. At the highest fields where all the moments are fully saturated, the magnetization should indeed be reduced by only 2.5%. It is interesting to speculate how this process takes place in these doped samples, and extremely high magnetic fields are required for such a study.

The onset fields of the magnetization plateaus, however, do not seem to shift with doping, suggesting a similar effective chemical potential of the triplet bosons for the pure and the doped samples. This is consistent with our simulation results which have found similar energy scales of the onset of the localized bound states $H'_{C2}$ in the pure and doped systems. Surprisingly, $H'_{C1}$ and $H'_{C3}$ are also relatively doping-insensitive within the doping range of our study. It is possible that higher doping concentration is required to observe their doping dependence.

In summary, our results provide a clear description of the magnetization process for $SrCu_{2-x}Mg_x(BO_3)_2$, revealing for the first time a rich impurity-induced phenomenology, suggesting that even for samples with a Mg-doping as low as 1% ~ 2.5%, a single-impurity description such as that discussed in Refs. 15,16 is not enough to capture the essential physics, and interactions between the impurities and triplets must be considered. Further studies with higher doping concentration are desired to better understand these impurity-induced emergent states, and to pursue the grand prize of RVB superconductivity.

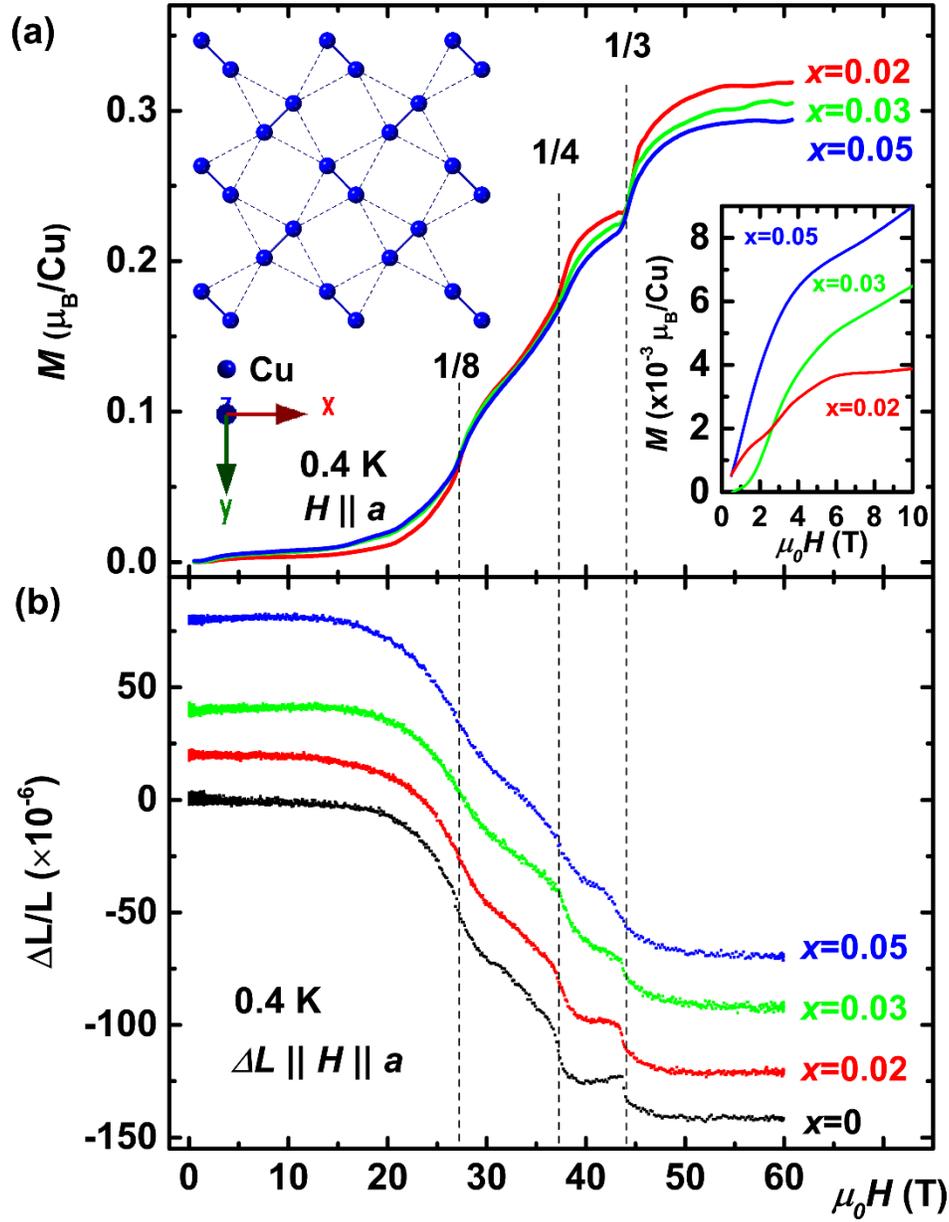

**Figure 1**: **Magnetization plateaus in Mg doped SCBO, with $H \parallel a$ axis, at $T = 0.4$ K. a**, Magnetization ($M$) and **b**, magnetostriction ($\Delta L/L$) vs. field, for $x = 0.02, 0.03$, and $0.05$, conducted in a 65 T multi-shot magnet at the pulsed field facility of the national high magnetic field laboratory (NHMFL). Lower right inset: $M(H)$ for the three dopings at low $H$. Data for $x = 0.02$ and $0.05$ is from 30 T shots, and data for $x = 0.03$ is from a 60 T shot. Upper left inset: a schematic of the spin-1/2 $Cu^{2+}$ atoms in the SSL lattice, as realized in $SrCu_2(BO_3)_2$. Traces in **b** are shifted for clarity.

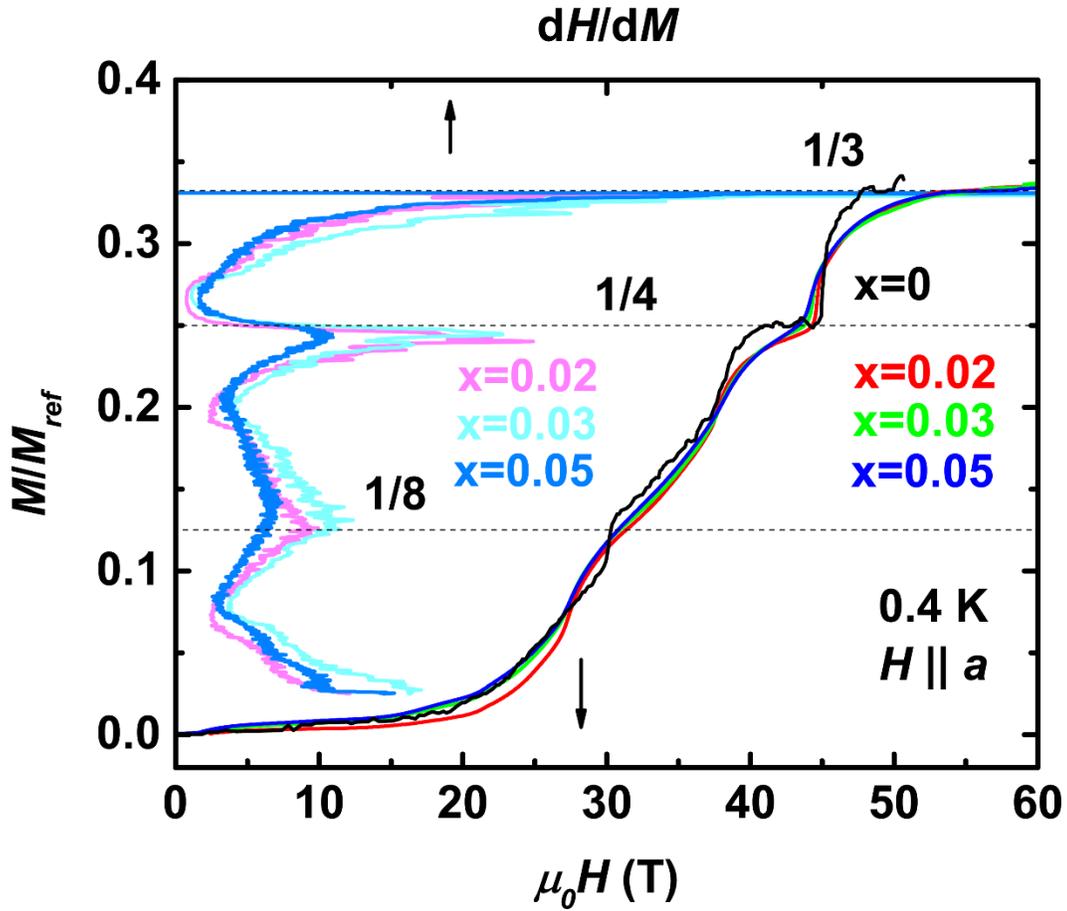

**Figure 2**: **Magnetization normalized by the reference saturation magnetization, $M_{ref}$ for each doping, with $H \parallel a$ axis, at $T = 0.4$ K.** Plateaus are indicated by peaks in the inverse susceptibility $dH/dM$ (top axis). Dashed lines guide the eye. The magnetization values at the 1/3 plateau are used to extract the reference saturation magnetization, $M_{ref}$ = 1.065, 0.952, 0.913, and 0.881 $\mu_B$/Cu for the $x = 0$, $x = 0.02$, $x = 0.03$, and $x = 0.05$ samples, respectively. Here, the $x = 0$ trace of $M$ vs. $\mu_0 H$ is reproduced from a $H \parallel c$ trace reported in ref. 24 (see Supplementary text), rescaled to allow comparison with our $H \parallel a$ data on the doped samples.

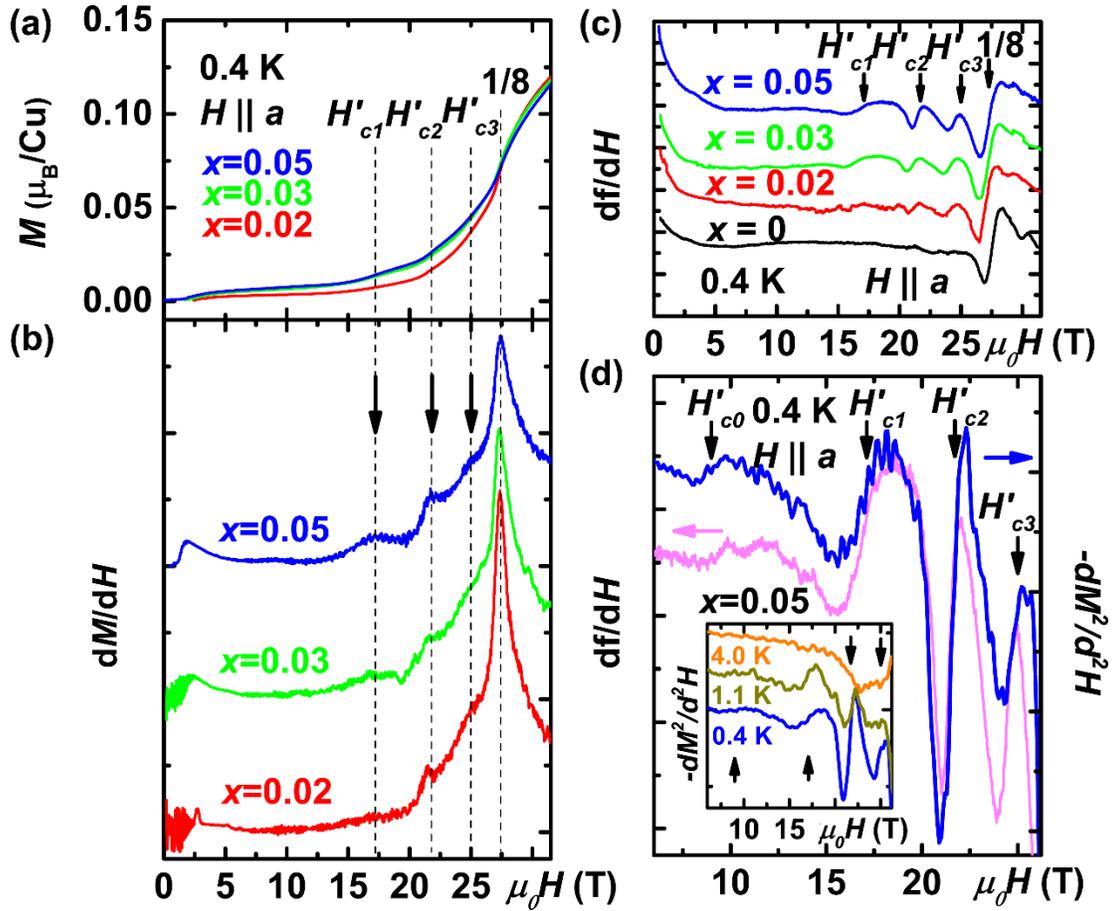

**Figure 3**: **Emergence of magnetization anomalies below the 1/8 plateau with $H \parallel a$ axis, at $T$ = 0.4 K.** Field dependence of **a**, $M$ and **b**, $dM/dH$, for three doping levels $x$ = 0.02, 0.03, and 0.05. The feature corresponding to $m = 1/8$ magnetization plateau is marked by a dashed line at ~ 27 T. $H'_{C1}$, $H'_{C2}$, and $H'_{C3}$ and the corresponding dashed lines indicate three of the additional anomalies below the 1/8 plateau. **c**, The first derivative of the frequency shift, $df/dH$ ($\propto dM^2/d^2H$), as a function of applied field for $x$ = 0, 0.02, 0.03, and 0.05, measured by tunnel diode oscillator technique at the dc field facility of NHMFL. Upon doping, three anomalies emerge at $H'_{C1}$, $H'_{C2}$, and $H'_{C3}$, consistent with the magnetization measurements in (**b**). **d**, A zoomed-in plot of $df/dH$ (left axis) and $-dM^2/d^2H$ (right axis) vs. $\mu_0 H$ for the $x$ = 0.05 sample, shows an additional broad anomaly at a lower field $H'_{C0}$. Inset: $-dM^2/d^2H$ vs. $\mu_0 H$ for a few selective temperatures. The arrows, from left to right, indicate $H'_{C0}$, $H'_{C1}$, $H'_{C2}$, and $H'_{C3}$, respectively. Traces in (**b**), (**c**), and inset of (**d**) are shifted for clarity.

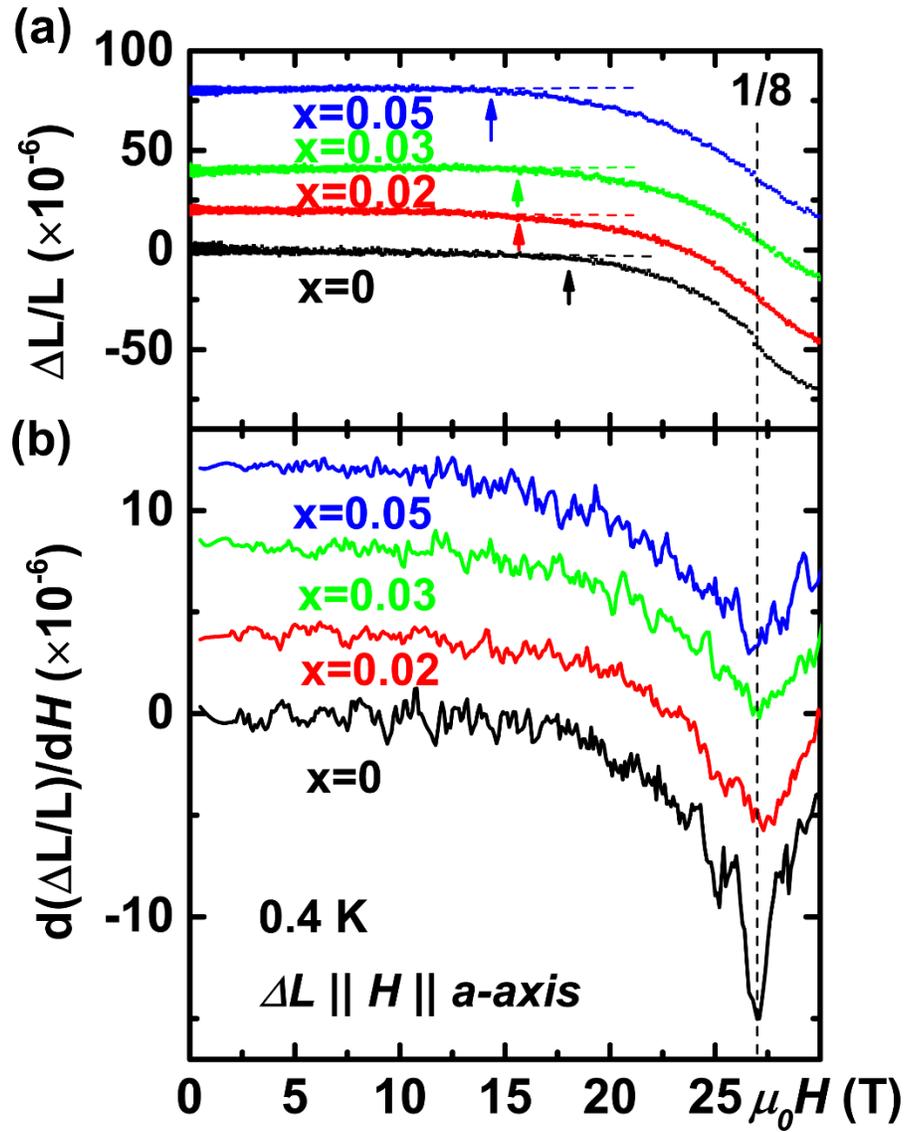

**Figure 4**: **Absence of the magnetization anomalies in magnetostriction measurements.** Field dependence of **a**, relative deformation $\Delta L/L$ along the tetragonal axis, and **b**, its first derivative $d(\Delta L/L)/dH$, with $\Delta L \parallel H \parallel a$ axis, at $T = 0.4$ K, for $x = 0, 0.02, 0.03,$ and $0.05$. The arrows in (**a**) point to the onsets of deviation from linear fits to the low field regions for each doping (dashed lines). The vertical dashed line marks the 1/8 plateau, identified as the local minima in $d(\Delta L/L)/dH$ vs. $\mu_0 H$. Traces are shifted for clarity.

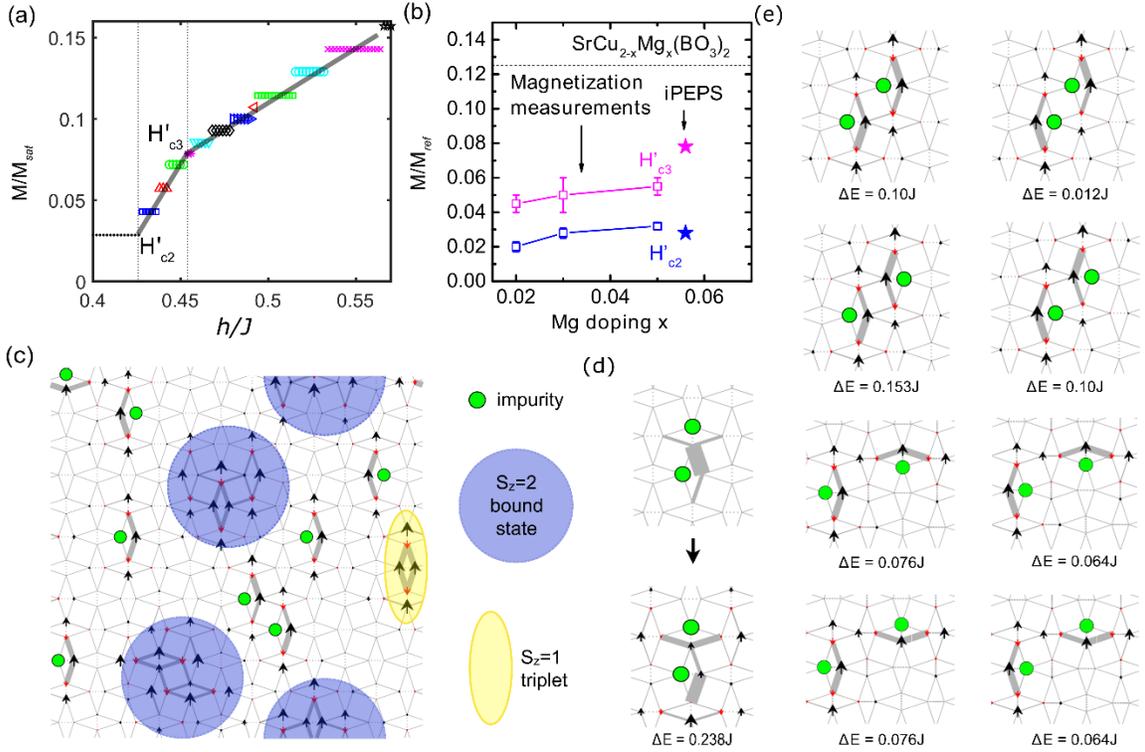

**Figure 5**: **iPEPS simulation results for the Mg doped SSL model, and its comparison with results of magnetization measurements. a**, iPEPS magnetization curve obtained using a 12 × 12 unit cell with 8 impurity sites. The full lines are guide to the eye. **b**, Mg-doping dependence of the normalized magnetization, $M/M_{ref}$, for experimental results on the $x$ = 0.02, 0.03, and 0.05 samples, and iPEPS simulation results for $M/M_{sat}$ with an effective doping $x$ = 0.056. Dashed line indicates the magnetization at the 1/8 plateau. **c**, Example spin configuration obtained in the total $S_z$ = 11 sector. The size of the spins scale with the magnitude of the local magnetic moment, where black (red) arrows point along (opposite to) the external magnetic field. The thickness of the grey bonds scales with the local bond energy (the thicker the lower the energy). **d**, Special 2-impurity configuration in the $S_z$ = 0 (top) and $S_z$ = 1 (bottom) sectors, respectively, with an excitation energy $\Delta E$ = 0.238$J$. **e**, Special 2-impurity configurations obtained with iPEPS which lead to additional excitation levels below $H'_{C1}$. Further distant impurities will have a smaller gap $\Delta E$, with $\Delta E \rightarrow 0$ in the limit of large separations. The excitation energies will be influenced also by additional impurities nearby. We note that the state with $\Delta E$ = 0.153$J$ is less relevant than the one at $H'_{C1}$ = 0.238J mentioned in the main text, since the probability of this configuration to appear is only half of the latter one.

# Methods

**Sample synthesis and characterization.** High quality single crystal samples of both SrCu$_2$(BO$_3$)$_2$ and SrCu$_{2-x}$Mg$_x$(BO$_3$)$_2$ ($x$ = 0.02, 0.03, and 0.05) were grown by the optical floating zone technique using self-flux, at a growth rate of 0.2 $mm\ h^{-1}$ in an $O_2$ atmosphere[52]. The $x$ = 0.02 and 0.03 samples were successfully grown for the first time, and were characterized by x-ray powder diffraction. The free $S$ = 1/2 impurities, i.e. the Mg-doping concentrations, for the $x$ = 0.02 and 0.03 samples were also characterized with measurements of the dc susceptibility as a function of temperature, using a commercial Quantum Design MPMS (see Fig. S1).

**Magnetization measurements.** Magnetization measurements were conducted on samples with approximate dimensions of ∼ 3.0 × 0.5 × 0.5 mm$^3$ ($a \times b \times c$) using a sample-extraction magnetometer in a 25 ms, 65 T pulsed magnet at the pulsed field facility of the National High Magnetic Field Laboratory in Los Alamos, NM[24]. The sample was placed inside a plastic capsule, which is inserted into or extracted from a pair of coaxial counterwound coils. The background signal was also determined for each temperature and subtracted from the data. Data was obtained for $H \parallel a$ down to 0.4 K, and calibrated with absolute values measured in a SQUID magnetometer from Quantum Design.

**Tunnel diode oscillator.** The TDO measurements[53] were carried out on cylindershaped crystals with approximate dimensions of ∼ 2 mm in length and ∼ 1 mm in diameter, at the dc field facility of the National High Magnetic Field Laboratory in Tallahassee, FL. A tunnel diode, operating in its negative resistance region, was used to provide power that maintains the resonance of a LC-circuit, at a frequency range between 10 and 50 MHz. The sample was placed inside a detection coil, with the a-axis of the sample aligned with the coil axis, forming the inductive component of the LC circuit. Changes in sample magnetization induce a change in the inductance, which is detected as a shift in the resonance frequency. The ability to measure the resonance frequency to a very high precision ensures the highly sensitive measurements in changes of magnetic moments ~ $10^{-12}$ *e.m.u.*.

**Torque magnetometry.** Torque magnetometry measurements were conducted to probe the susceptibility anisotropy of samples with approximate dimensions of ~ 0.2 × 0.2 × 0.2 mm$^3$ ($a \times b \times c$) in static magnetic field[22]. Samples were attached with silicone grease to the commercial piezoresistive atomic force microscopy (AFM) cantilevers (Seiko PRC400) (Ref. 54), which form a Wheatstone bridge configuration with two additional adjustable resistors. Changes of sample magnetization with field induced torque on the cantilever, and are detected as a voltage across the bridge.

**Magnetostriction measurements.** An optical fiber, equipped with a 1–*mm*-long fiber Bragg grating (FBG), was attached to single crystal samples with approximate dimensions of ~ 3.0×0.5×0.5 mm$^3$ ($a \times b \times c$) along their a-axes, using cyanoacrylate. The samples were held in place solely by the fiber, and were orientated such that the applied field is parallel with the a-axes of the samples. The FBG is illuminated by a broadband light (1525 - 1565 nm) source, and reflects a narrow band of light (≈ 1550 nm) (Ref. 55). The length variation $\Delta L/L$ along a-axis axial configuration is detected by monitoring the shift of the reflected light by the FBG[20].

## Acknowledgements

We are grateful to V. Zapf for help in acquiring the PPMS/MPMS data. We thank T.F. Rosenbaum, D.M. Silevitch, C.D. Batista and B.D. Gaulin for helpful discussions. A portion of this work was performed at the National High Magnetic Field Laboratory, which is supported by the National Science Foundation Cooperative Agreement No. DMR1157490 and DMR-1644779, the State of Florida and the U.S. Department of Energy. We acknowledge the use of the Analytical Instrumentation Facility (AIF) at North Carolina State University, which is supported by the State of North Carolina and the National Science Foundation (award number ECCS-1542015). Z.S., W.S., C.M., and S.H. acknowledge support provided by funding from the Powe Junior Faculty Enhancement Award, and William M. Fairbank Chair in Physics at Duke University. P.C. and F.M. acknowledge the support provided by Swiss National Science Foundation and the European Research Council (ERC) under the European Union's Horizon 2020 research and innovation programme (grant agreement No 677061).

## Author contributions

Research conceived and supervised by S.H.; Single-crystal SCBO samples grown by C.M, H.A.D. and S.H.; High-field measurements performed by Z.S., W.S., D.G., F.W., N.H., M.J. and S.H., and analyzed by Z.S., W.S. and S.H.; iPEPS calculations performed by P.C. and F.M.; Manuscript written by Z.S., P.C. and S.H.; all authors commented on the manuscript.

## Additional information

Supplementary information accompanies this paper.

## Competing financial interests

The authors declare no competing financial interests.

# Supplementary Material for: Emergent Bound States and Impurity Pairs in Chemically Doped Shastry-Sutherland System

**Supplementary Text**

**Data reproduced from previous magnetization measurements on the pure sample**

The magnetization measurements in Ref. 1 were performed on the pure ($x = 0$) sample with $H||c$, and a background $0.14\times10^3$ *emu/mol Cu*, which was attributed to crystalline defects, was subtracted. In our Mg-doped samples, however, impurities due to crystalline defects could not be differentiated from the Mg-induced spin impurities. Therefore, the background was added back to the reproduced $x = 0$ data from Ref. 1, to allow a fair comparison. Moreover, in the reproduced $x = 0$ data, $\mu_0 H$ is multiplied by the *g*-factor ratio $g_{||c}/g_{||a}$ = 2.28/2.04 = 1.12 (Ref. 2), and *M* is divided by the same *g*-factor ratio, to account for the different field orientations.

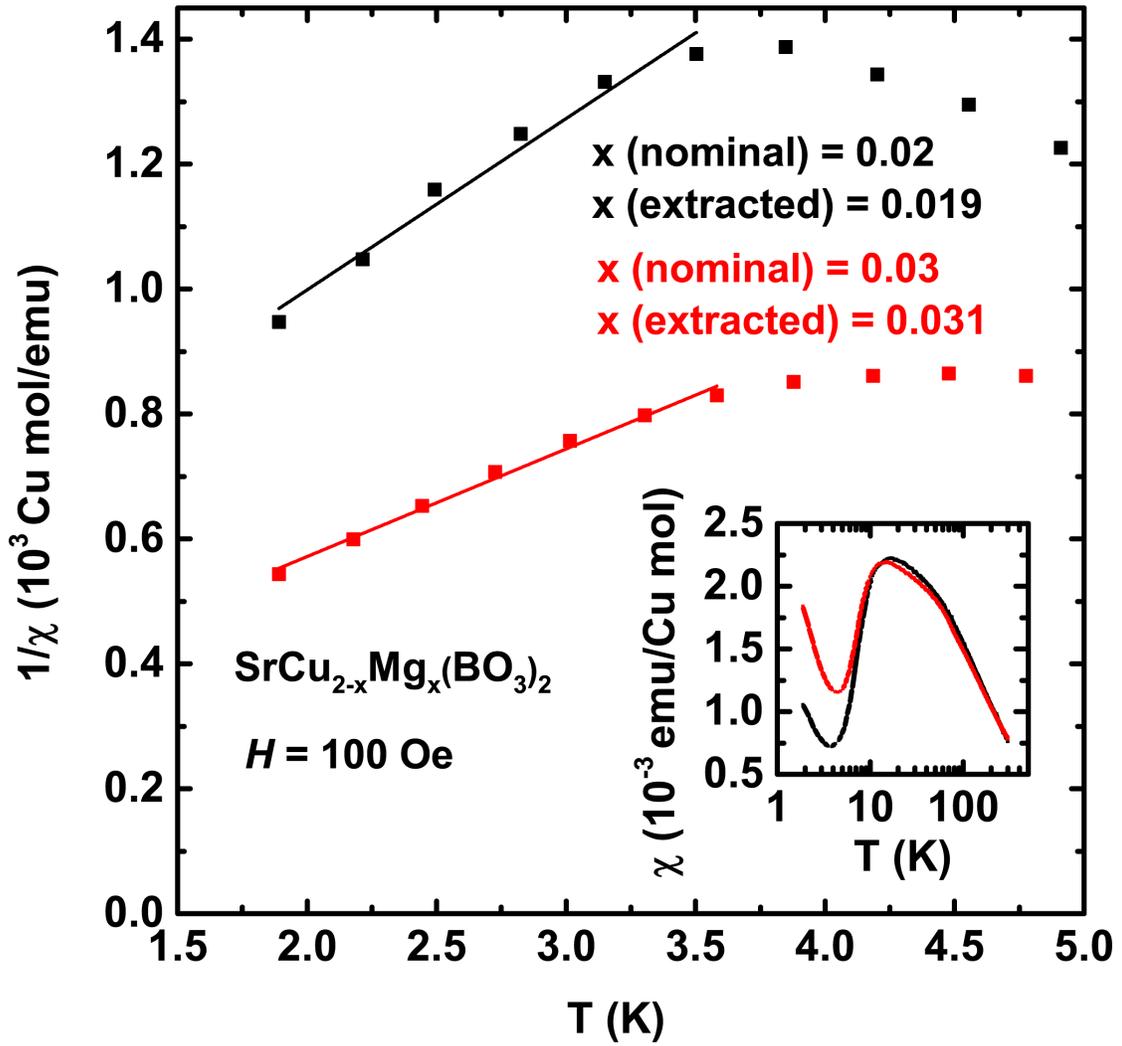

**Figure S1**: **Temperature dependence of the inverse magnetic susceptibility for the x = 0.02, 0.03 samples in an applied field of 100 Oe, parallel to the *ab* plane.** The fits to the Curie-Weiss law $C'/(T - \theta')$ in the 1.8 K - 3.5 K regime, as shown by the solid lines, give $C'(lowT)$ = $3.6 \times 10^{-3}$ *emu K/Cu mol* and $5.8 \times 10^{-3}$ *emu K/Cu mol* for the $x = 0.02$ and $x = 0.03$ samples respectively, which correspond to $x = 0.019$ and $x = 0.031$, assuming free $S = 1/2$ impurity spins. (Inset) $\chi(T)$ for the entire temperature range (1.8 K ≤ $T$ ≤ 300 K).

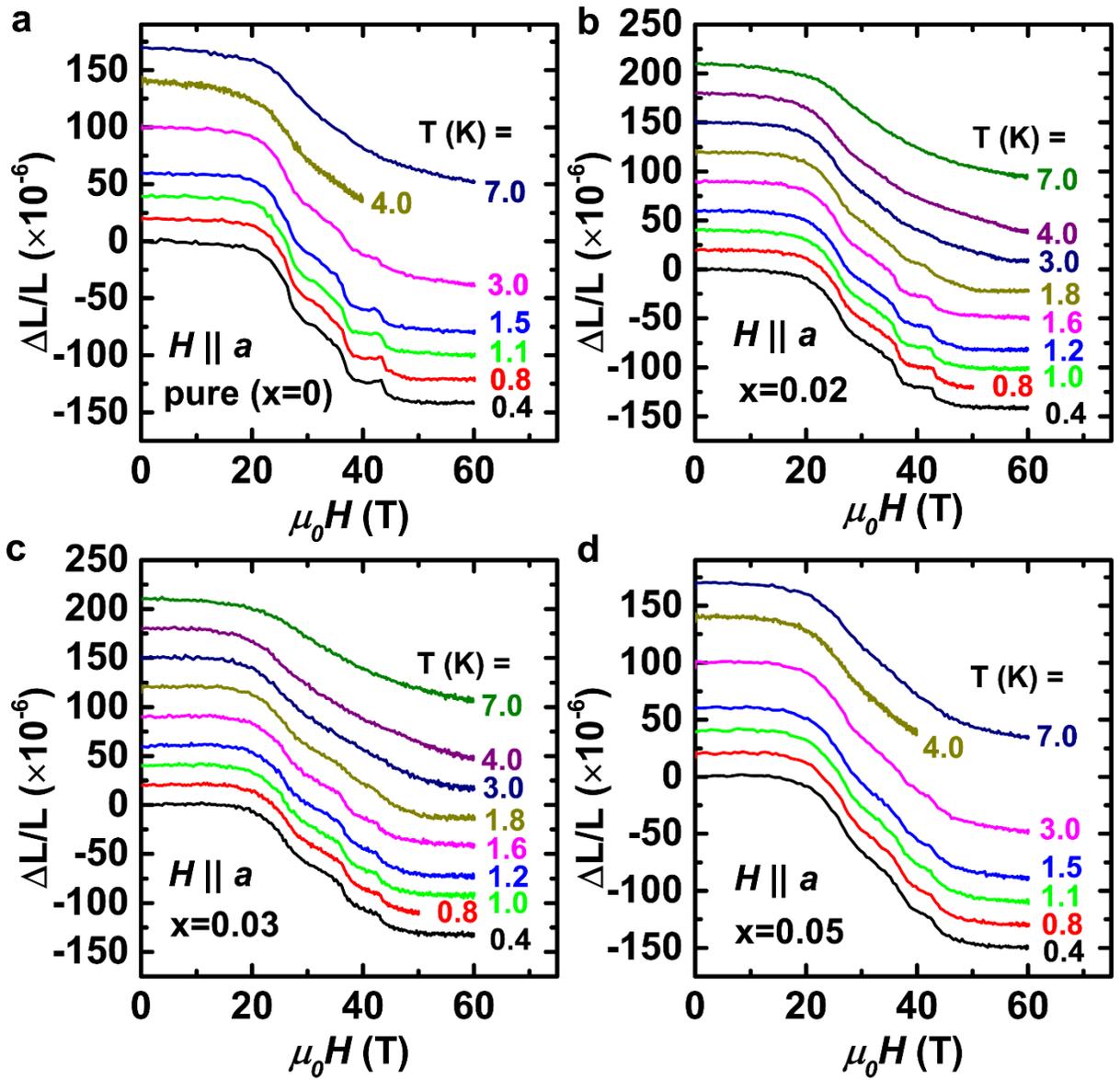

**Figure S2**: **Temperature evolution of the plateaus for (a) the pure ($x$ = 0) and (b – d) the Mg-doped ($x$ = 0.02, 0.03, 0.05) samples in magnetostriction measurements.** Field ($H \parallel a$ axis) dependence of $\Delta L/L$ in pulsed fields up to 60 T. Data presented was taken during field upsweep. Traces are shifted for clarity.

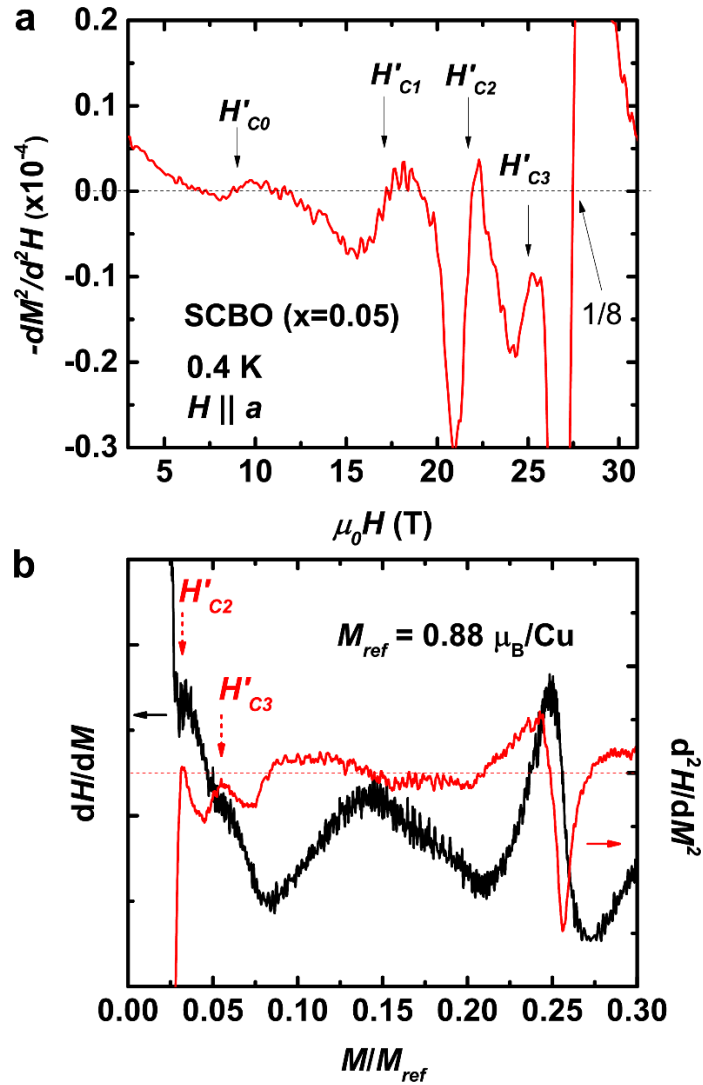

**Figure S3**: **Method to determine the onset magnetic fields and magnetization values of the anomalies for the** $x = 0.05$ **sample at** $T = 0.4$ **K. a**, $H'_{C0}$, $H'_{C1}$, and $H'_{C2}$ are defined as the fields where $-d^2M/dH^2$ crosses zero from below, i.e., peaks in $dM/dH$ vs. $H$. $H'_{C3}$ is defined as the peak in $-d^2M/dH^2$ vs. $H$. **b**, The inverse susceptibility $dH/dM$ (black curve, left axis) and the second derivative $d^2H/dM^2$ (red curve, right axis) vs. $M/M_{ref}$. The magnetization values $M/M_{ref}$ at $H'_{C2}$ and $H'_{C3}$ are determined as the maxima in the second derivative, as shown with the red dashed arrows. The analyses were repeated for the $x = 0.02$ and $0.03$ samples.

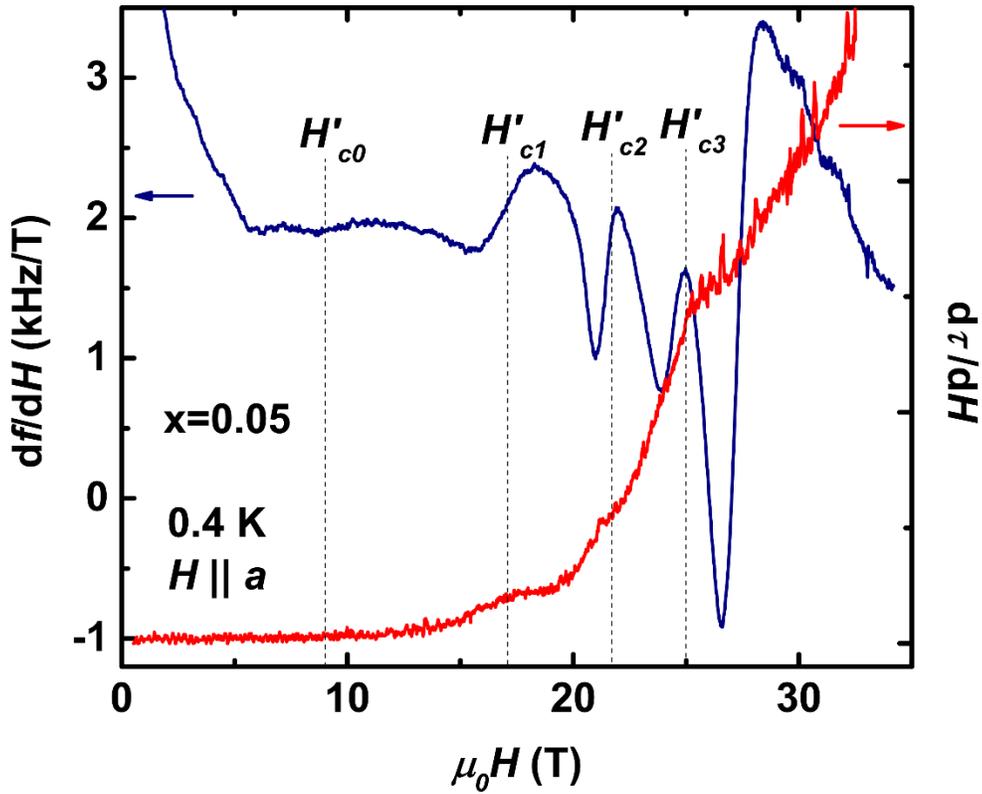

**Figure S4**: **The TDO and torque magnetometry measurements for the $x$ = 0.05 sample.** Field ($H \parallel a$ axis) dependence of $df/dH$ ($\propto dM^2/d^2H$) from the TDO measurements (blue, left axis), and $d\tau/dH$ from the torque magnetometry measurements (red, right axis) at 0.4 K. The $H'_{C0}$, $H'_{C1}$, $H'_{C2}$, and $H'_{C3}$ anomalies, identified from the TDO and magnetization measurements are indicated by the dashed lines. The latter three anomalies also appear in the torque magnetometry measurements. Measurements were repeated for the $x$ = 0.02 and $x$ = 0.03 samples, and similar results were obtained.

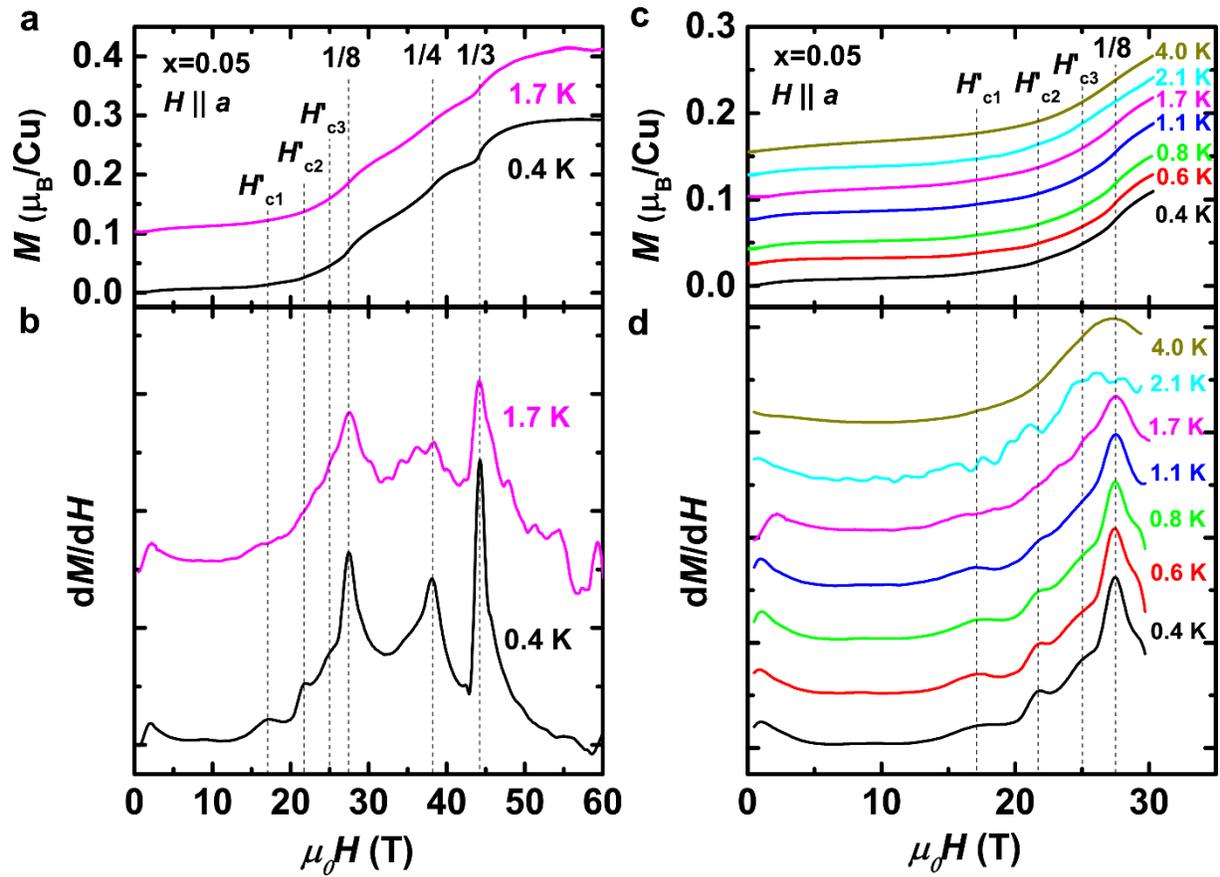

**Figure S5**: **Evolution of the $H'_{C1}$, $H'_{C2}$ and $H'_{C3}$ anomalies and the plateaus with temperature for the x = 0.05 sample.** Field ($H \parallel a$ axis) dependence of (**a** and **c**) $M$ and (**b** and **d**) $dM/dH$ in pulsed fields up to 60 T and 30 T. The $H'_{C1}$, $H'_{C2}$ and $H'_{C3}$ anomalies and the plateaus are indicated by the dashed lines. Data presented was taken during field upsweep. Traces are shifted for clarity.

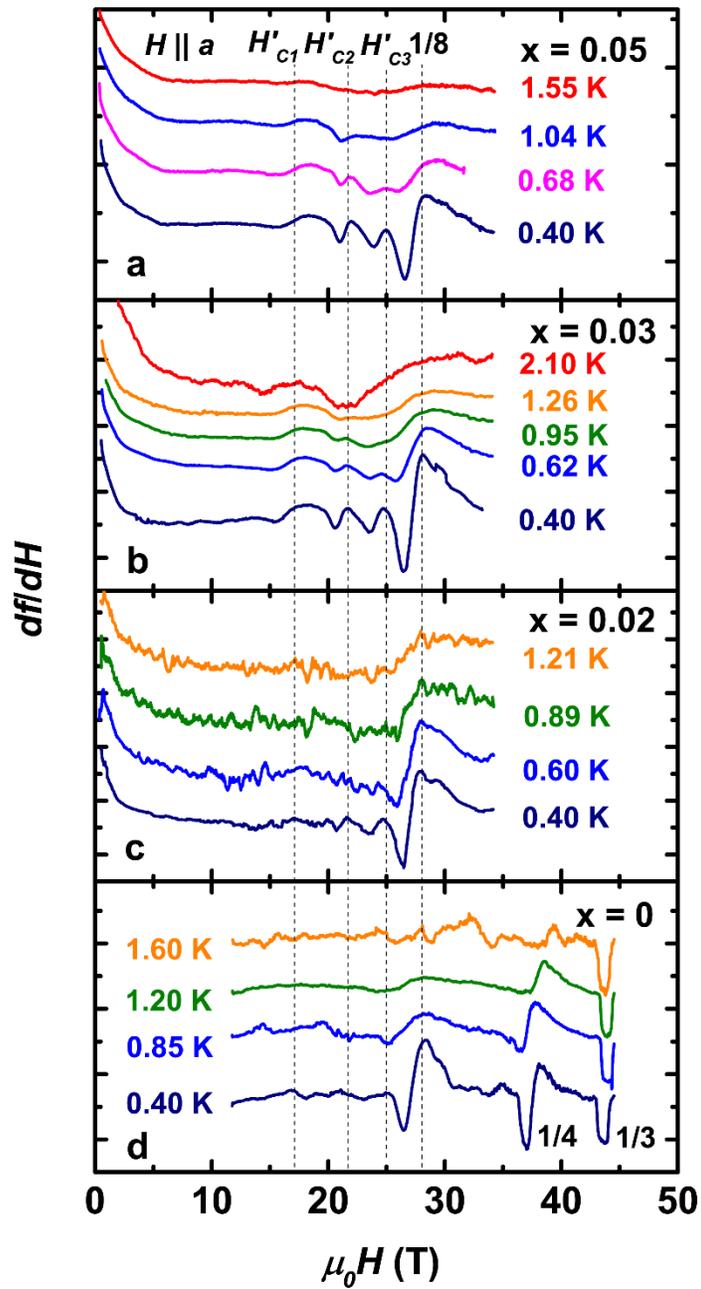

**Figure S6**: **Evolution of the $H'_{C1}$, $H'_{C2}$ and $H'_{C3}$ anomalies and the 1/8 plateau with temperature for (a – c) the Mg-doped ($x = 0.05, 0.03, 0.02$) and (d) the pure ($x = 0$) samples in the TDO measurements.** Field ($H \parallel a$ axis) dependence of $df/dH$ ($\propto dM^2/d^2H$) in static fields. Traces are shifted for clarity. Dashed lines guide the eye.

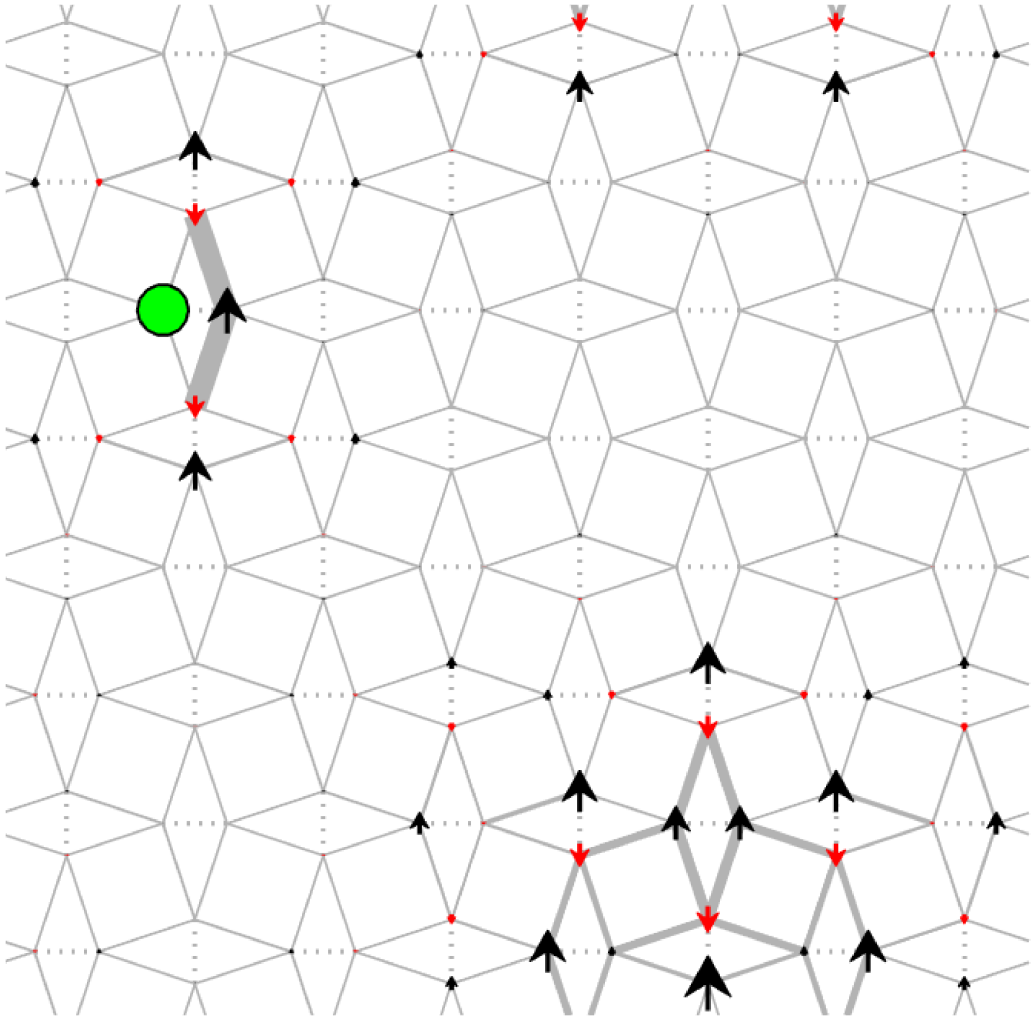

**Figure S7**: **iPEPS results for lowest energy state in an 8×8 unit cell.** Simulations periodically repeated in the infinite lattice; including one impurity (green disc) with a neighboring S=1/2 site and one $S_z$ = 2 bound state (partially delocalized) away from the impurity, showing that a bound state is not attracted but repelled by an impurity site. The size of the spins scale with the magnitude of the local magnetic moment, where black (red) arrows point along (opposite to) the external magnetic field. The thickness of the grey bonds scales with the local bond energy (the thicker the lower the energy).